\documentclass[cernpreprint=true, atlasdraft=false, USenglish, orcidlogo, texmf, orcidlogo]{atlasdoc}
\usepackage[subfigure,rotating]{atlaspackage}
\usepackage{atlasbiblatex}
\usepackage{multirow}

\usepackage{atlasphysics}
\graphicspath{{logos/}{figures/}}

\usepackage{ANA-HDBS-2023-19-PAPER-defs}

\AtlasTitle{Combination of searches for singly and doubly charged Higgs bosons produced via vector-boson fusion in proton--proton collisions at $\sqrt{s}=13$ TeV with the ATLAS detector}

\AtlasAbstract{A combination of searches for singly and doubly charged Higgs bosons, $H^{\pm}$ and $H^{\pm\pm}$, produced via vector-boson fusion is performed using 140~fb$^{-1}$ of proton--proton collisions at a centre-of-mass energy of 13 TeV, collected with the ATLAS detector during Run 2 of the Large Hadron Collider. Searches targeting decays to massive vector bosons in leptonic final states (electrons or muons) are considered. New constraints are reported on the production cross-section times branching fraction for charged Higgs boson masses between 200 GeV and 3000 GeV. The results are interpreted in the context of the Georgi-Machacek model for which the most stringent constraints to date are set for the masses considered in the combination.}

\AtlasRefCode{HDBS-2023-19}

\PreprintIdNumber{CERN-EP-2024-189}

\arXivId{2407.10798}

\HepDataRecord{https://doi.org/10.17182/hepdata.153847.v2}

\AtlasJournal{Phys. Lett. B}
\AtlasJournalRef{Phys. Lett. B 860 (2025) 139137}
\AtlasDOI{10.1016/j.physletb.2024.139137}

\hypersetup{pdftitle={ATLAS document},pdfauthor={The ATLAS Collaboration}}

\begin{document}

\maketitle

\tableofcontents

\section{Introduction}
\label{sec:intro}

An important avenue of research for physics beyond the Standard Model (SM) is to understand whether the Higgs boson discovered in 2012~\cite{ATLAS:2012yve, CMS:2012qbp, HIGG-2021-23, CMS-HIG-22-001} at the Large Hadron Collider (LHC) is part of an extended Higgs sector. Charged Higgs bosons are predicted in extended Higgs sectors with additional complex doublets~\cite{PhysRevD.8.1226, FAYET197414} or with additional higher-isospin scalar fields~\cite{Mohapatra:1974hk,Arkani-Hamed:2002ikv,Han:2003wu}. In the generic two Higgs doublet model the charged Higgs boson $H^{\pm}$ does not decay into $WZ$ bosons as a result of CP-invariance which forbids a tree-level $H^{\pm}W^{\pm}Z$ coupling. A tree-level coupling to massive vector bosons, is, however, present in models with additional isotriplet scalar fields~\cite{Englert:2013wga}.

The Georgi--Machacek (GM) model~\cite{Georgi:1985nv,Chanowitz:1985ug} is used as a benchmark in this Letter. The GM model extends the Higgs sector of the SM by including one real and one complex triplet. This preserves a custodial symmetry at tree level whereby the GM model is not strongly constrained~\cite{Hartling:2014aga}. A parameter, $\sin \theta_H $, characterises the contribution of the isotriplet scalar fields to the masses of the $W$ and $Z$ bosons. The physical scalar states are organised into distinct custodial multiplets: a quintuplet $(H_5^{\pm\pm},H_5^{\pm},H_5^0)$ that is fermiophobic but couples to $W$ and $Z$ bosons, a triplet, and two singlets, one of which is identified as the observed $125~\GeV$ SM-like Higgs boson. The production cross-sections and the widths of the $H_5^{\pm\pm}$ and $H_5^{\pm}$ states are proportional to $\sin^2 \theta_H$. At tree level, the physical states in each multiplet are degenerate in mass, denoted as $m_{H_{5}}$ for the quintuplet. Differences in mass of up to a few~\GeV in the quintuplet due to higher order effects~\cite{Keeshan:2018ypw} are not considered in this analysis.

\begin{figure}[t!bp]
\centering
\subfigure[]{\includegraphics[width=0.40\textwidth]{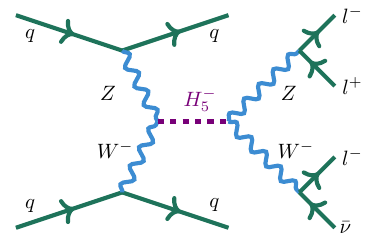}}
\subfigure[]{\includegraphics[width=0.40\textwidth]{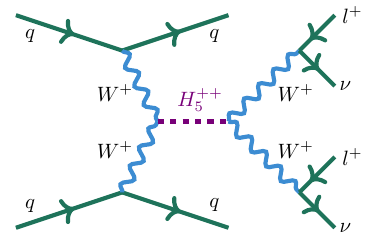}}
\caption{Representative Feynman diagrams of production and decay of the (a) $H_5^{\pm}$ and (b) $H_5^{\pm\pm}$ states.}
\label{fig:feyn_sig}
\end{figure}

The production of the $H_5^{\pm}$ and $H_5^{\pm\pm}$ scalars is mediated via vector-boson fusion (VBF) in the GM model. Figure~\ref{fig:feyn_sig} shows representative Feynman diagrams of production and decay of the $H_5^{\pm}$ and $H_5^{\pm\pm}$ states. The \texttt{H5plane} benchmark is considered~\cite{deFlorian:2016spz}, where the triplet states are assumed to be heavier than the quintuplet states. Thus, in this benchmark, the branching fraction of $\hp$ and $\hpp$ decays is 100\%~\cite{LHCHXSWG-2015-001}. For the parameter space explored in this combination, the considered intrinsic width of the $H_5^{\pm\pm}$ and $H_5^{\pm}$ states is below 5\%, which is smaller than the experimental resolution.

This Letter reports the combination of the ATLAS Collaboration searches for $\hp$~\cite{atlas:2023wz} and $\hpp$~\cite{ATLAS:2023sua} produced via VBF using proton--proton (\pp) collisions at $\sqrt{s}=13~\TeV$. The dataset corresponds to an integrated luminosity of $140.1 \pm 1.2$~fb$^{-1}$~\cite{DAPR-2021-01,LUCID2}, collected with the ATLAS detector~\cite{PERF-2007-01} during Run 2 of the LHC (2015--2018). An extensive software suite \cite{SOFT-2022-02} is used in data simulation, in the reconstruction and analysis of real and simulated data, in detector operations, and in the trigger and data acquisition systems of the experiment. The searches target events where the vector boson decays include electrons or muons.  Model independent upper limits at 95\% confidence level (CL) are reported on the production cross-section times branching fraction for the VBF production of the $H_5^{\pm}$ and $H_5^{\pm\pm}$ bosons individually. The results are interpreted in the context of the GM model, including the simultaneous contributions of $H_5^{\pm}$ and $H_5^{\pm\pm}$, setting upper limits on the $\sin \theta_H$ parameter as a function of $m_{H_{5}}$. A simultaneous search for the $H_5^{\pm}$ and $H_5^{\pm\pm}$ states in the VBF topology using the same leptonic decay modes and $137$~fb$^{-1}$ of integrated luminosity was published by the CMS Collaboration~\cite{CMS-HIG-20-017}, reporting upper bounds at 95\% CL on the $\sin\theta_H$ parameter. The reported CMS upper bounds vary between $\sim 0.2$ and 0.55 in the mass range 200--2000~\GeV.

Constraints on the GM model have also been reported by the ATLAS Collaboration considering the VBF production of the $H_{5}^0$ state and the $W^{\pm}W^{\mp}$ boson pair decay channel~\cite{HIGG-2016-31} at $\sqrt{s}=13~\TeV$ using an integrated luminosity of $36.1$~fb$^{-1}$. In addition, the ATLAS and CMS Collaborations have reported constraints on the GM model at $\sqrt{s}=8~\TeV$~\cite{ATLAS:2015edr} and $\sqrt{s}=13~\TeV$ with $35.9$~fb$^{-1}$ of integrated luminosity \cite{CMS:2019qfk}, respectively, via searches for $\hp$ and $\hpp$ using semileptonic final states. The results from Refs.~\cite{ATLAS:2015edr, HIGG-2016-31} are not included in this combination due to their limited sensitivity. Constraints from searches for heavy neutral Higgs bosons produced via VBF and decaying into $WW$ or $ZZ$ channels can also be interpreted in the GM model as discussed in Ref.~\cite{ATL-PHYS-PUB-2022-008}, but a combination of all these searches is beyond the scope of this Letter.

\section{Description of the nominal analyses}
\label{sec:analysis}

A brief description of the $\hp$ and $\hpp$ analyses is given below. The detailed information about the reconstruction, identification, and calibration of physics objects, as well as the simulation, triggers, and event selection used in these results, can be found in Refs.~\cite{atlas:2023wz,ATLAS:2023sua}, which form the basis for this combination analysis. The final states probed by these analyses consist in the leptonic decays (electrons or muons) of two massive vector bosons produced in association with two jets. The event selection in the $H^{\pm}$ signal region (SR) requires three charged leptons, while the $H^{\pm\pm}$ SR requires a same-charge lepton pair. In the $H^{\pm}$ SR, if there is more than one pair of same flavour and opposite charge leptons that can form a $Z$ boson candidate, the one with invariant mass closest to the $Z$ boson pole mass is chosen. The third lepton is then taken as the $W$ boson lepton candidate. The missing transverse momentum, with magnitude \MET, is required to be larger than 25 (30)~\GeV in the $H^{\pm}$ ($H^{\pm\pm}$) SR to exploit the presence of neutrinos in the final state. The VBF topology is characterised by requiring at least two jets with a large invariant mass, $\mjj$, and a large absolute rapidity difference, $\left|\dyjj\right|$. The invariant mass of the two highest transverse momentum jets must satisfy $\mjj>100~\GeV$ in the $H^{\pm}$ SR and is further constrained as discussed below. The $H^{\pm\pm}$ SR is defined by requiring $\mjj>500~\GeV$ and $\left|\dyjj\right|>2.0$.

The presence of only one neutrino in the $\hp$ channel is exploited to estimate the longitudinal component of the neutrino momentum by constraining the invariant mass of the charged lepton and neutrino system to the pole mass of the $W$ boson, where the charged lepton is the one assigned to the $W$ boson candidate. The \MET is assumed to be due to the neutrino. The resulting quadratic equation leads to two solutions. If they are real, the one with the smaller magnitude of the neutrino momentum is chosen, otherwise, the real part is chosen. The choice of the solution was optimised using generator-level information. The resulting reconstructed $W$ boson four-momentum is used to calculate the invariant mass of the $WZ$ system, $m_{WZ}$, which is used as a discriminating variable between the resonant $H^{\pm}$ signal and the SM backgrounds. The full kinematic reconstruction of the invariant mass of the $W^{\pm}W^{\pm}$ system is not attempted in the $H^{\pm\pm}$ SR due to the presence of two neutrinos. The transverse mass, \mt, defined as
\begin{equation*}
\mt = \sqrt{\left(\ET^{\ell\ell} + \MET \right)^2 - \left|\vec{p}_\mathrm{T}^{\ell\ell} + \vec{E}^{\mathrm{miss}}_{\mathrm{T}} \right|^2},
\end{equation*}
where \( \ET^{\ell\ell} \) is the transverse energy of the dilepton system, \( \vec{p}_\mathrm{T}^{\ell\ell} \)
is the vectorial sum of the lepton transverse momenta, and \( \vec{E}^{\mathrm{miss}}_{\mathrm{T}} \) is the missing transverse momentum vector, is used as a discriminating variable in the $H^{\pm\pm}$ SR.

\begin{figure}[t!bp]
\centering
\subfigure[]{\includegraphics[width=.24\textwidth]{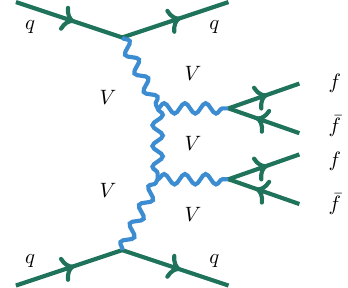}}
\subfigure[]{\includegraphics[width=.24\textwidth]{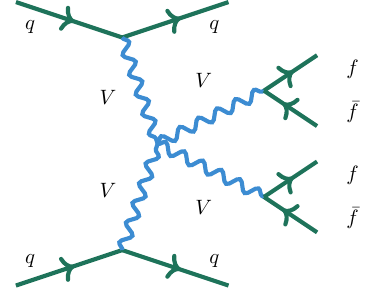}}
\subfigure[]{\includegraphics[width=.24\textwidth]{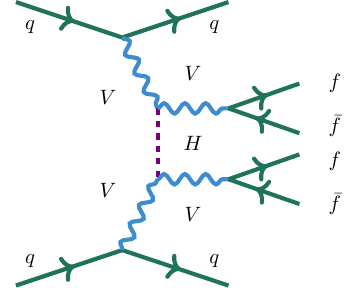}}
\caption{Representative Feynman diagrams for a VBS EW $VVjj$ production that either include (a) a triple-gauge-boson vertex, (b) a quartic gauge boson vertex, and (c) the exchange of a Higgs boson in the $t$-channel. The lines are labelled by quarks ($q$), vector bosons ($V=W/Z$), the Higgs boson ($H$) and fermions ($f$).}
\label{fig:diagrams_ewk_vbs}
\end{figure}

\begin{figure}[t!bp]
\centering
\includegraphics[width=.24\textwidth]{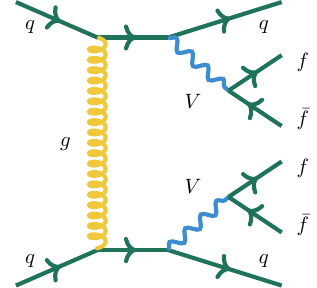}\hfill
\includegraphics[width=.24\textwidth]{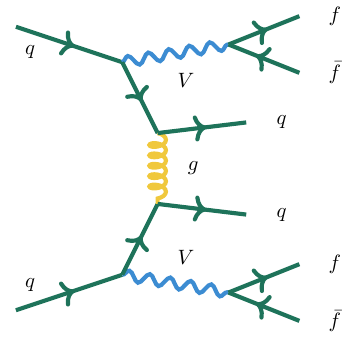}\hfill
\includegraphics[width=.24\textwidth]{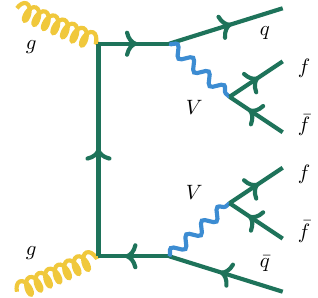}\hfill
\includegraphics[width=.24\textwidth]{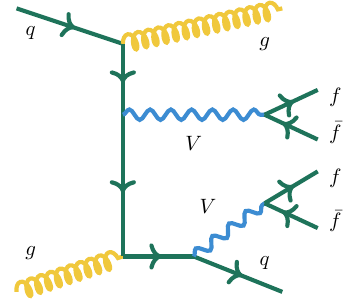}\\
\caption{Representative Feynman diagrams for QCD $VVjj$ production with strong interaction vertices. The lines are labelled by quarks ($q$), vector bosons ($V=W/Z$), fermions ($f$), and gluons (g). The two diagrams on the right with gluons in the initial state are not present for \sswwjj production.}
\label{fig:diagrams_qcd}
\end{figure}

The SM production of two vector bosons in association with two jets, denoted as $VVjj$, constitutes the dominant background for these searches. The production of $VVjj$ at leading-order (LO) has contributions both from processes that involve only electroweak (EW) interaction vertices (EW $VVjj$) and from processes that involve strong interaction vertices (QCD $VVjj$). Representative Feynman diagrams for the vector-boson scattering (VBS) processes, which are part of a class of processes contributing to the EW $VVjj$ production, are shown in Figure~\ref{fig:diagrams_ewk_vbs}. Representative Feynman diagrams for the QCD $VVjj$ processes are shown in Figure~\ref{fig:diagrams_qcd}. The QCD and EW $\wzjj$, and EW $\sswwjj$ processes are the dominant backgrounds in the $H^{\pm}$ and $H^{\pm\pm}$ SRs, respectively. The QCD and EW $\wzjj$ processes are important backgrounds also in the $H^{\pm\pm}$ SR, contributing 22\% of the overall expected event yield. It contributes when one of the leptons is not selected, typically because it is outside of the geometrical acceptance of the detector.

The dominant background processes are estimated with Monte Carlo (MC) simulated events and their modelling is constrained in dedicated signal-depleted control regions (CRs). An artificial neural network (ANN) with a binary classification task is used for the $H^{\pm}$ search to categorise the events as belonging to either the VBF process or to a background process. The ANN is trained with simulated $H^{\pm}$ events as signal, against QCD and EW $\wzjj$ events as background. The $H^{\pm}$ SR is defined by requiring the ANN score to be greater than 0.82, which maximises the significance and effectively starts the SR at $\mjj>500~\GeV$. Events with an ANN score of less than 0.82 and $\mjj>500~\GeV$ are used to define the QCD $\wzjj$ CR.

Events with $200 < \mjj < 500~\GeV$ are used in the $H^{\pm\pm}$ search to define the ``low-\mjj'' CR. This CR has a similar background composition to the SR and is used to control the uncertainties of major background contributions. A dedicated QCD $\wzjj$ CR is defined by requiring events with three charged leptons in the final state, $\mjj>200~\GeV$, and a trilepton invariant mass greater than 106~\GeV.

Leptons from hadron decays and jets misidentified as leptons are referred to as non-prompt leptons. The non-prompt lepton background is the third-largest background process in the $H^{\pm\pm}$ SR and arises mainly from $W$+jets and semileptonic $t\bar{t}$ processes. The non-prompt lepton, electron charge misidentification and photon conversion backgrounds are estimated using data-driven methods as described in Ref.~\cite{ATLAS:2023sua}, and contribute 17\% of the overall expected event yield. The contribution of the $ZZ$ process is a non-negligible background in the $H^{\pm}$ SR. By requiring four leptons in the final state and removing the \MET requirement a corresponding $ZZ$ CR is defined to extract the normalisation. Small background contributions from triboson $VVV$ and $t\bar{t}V$ are included in both the $H^{\pm}$ and $H^{\pm\pm}$ SRs, while the $tZq$ process is considered only in the $H^{\pm\pm}$ SR.

\section{Combination strategy}
\label{sec:stat}

A simultaneous binned maximum-likelihood fit is performed to determine the $\hp$ and $\hpp$ signal strength. The $m_{WZ}$ distribution is fitted in the $H^{\pm}$ SR, while the \mt distribution in five regions of $\mjj$ with boundaries at (500, 850, 1450, 2100, 2550, $\infty$)~\GeV is fitted in the $H^{\pm\pm}$ SR. The SRs and the $ZZ$, low-\mjj, and QCD $\wzjj$ CRs are fitted simultaneously. The QCD $\wzjj$ CR defined in Ref.~\cite{ATLAS:2023sua} overlaps considerably with the QCD $\wzjj$ CR defined in Ref.~\cite{atlas:2023wz} and is not strictly orthogonal to the $H^{\pm}$ SR. Consequently, only the QCD $\wzjj$ CR defined in Ref.~\cite{atlas:2023wz} is used in this simultaneous fit. The resulting impact on the $\hpp$ signal extraction when either of the QCD $\wzjj$ CRs is used was found to be negligible.

The signal, EW $\sswwjj$, QCD $\wzjj$, and $ZZ$ background normalisations are kept as floating parameters in the fit and are constrained by the data in both the SRs and dedicated CRs. The relatively small contributions of the EW $\wzjj$ and QCD $\sswwjj$ processes are normalised to the SM predictions and allowed to vary within their uncertainties. The systematic uncertainties are included as nuisance parameters~\cite{Rolke:2004mj} with Gaussian priors. The nuisance parameters are adjusted in the fit with the shape and normalisation of each distribution varying within the specified constraints. The results are driven by the statistical uncertainty of the data in the SRs and none of the considered systematic uncertainties have significant impact on the sensitivity of this search. The expected limits on $\sin \theta_H $ improve by up to 5\%, depending on $m_{H_{5}}$, if the systematic uncertainties are not included in the simultaneous fit. The largest systematic uncertainties considered are briefly discussed in the following.

The dominant contributions to the systematic uncertainties stem from the theoretical uncertainties in the physics modelling of the GM signal, followed by the experimental uncertainty sources related to the jet energy calibration~\cite{JETM-2018-05}. The $H^{\pm}$ ($H^{\pm\pm}$) signal samples are simulated with \MGNLO[2.7.2 (2.9.5)]~\cite{Alwall:2014hca} interfaced to \PYTHIA{8.186 (8.245)}~\cite{Sjostrand:2007gs,Sjostrand:2014zea} for the modelling of the parton shower in the ``dipole recoil'' scheme~\cite{Sjostrand:2018ds}, hadronisation and underlying event. The $H^{\pm}$ and $H^{\pm\pm}$ samples are simulated at next-to-LO (NLO) and LO in QCD, respectively. For a given $H^{\pm}$ or $H^{\pm\pm}$ sample, no significant differences in the shapes between the LO and NLO distributions are seen. When deriving the constraints on $\sin \theta_H$, the next-to-NLO (NNLO) predictions~\cite{LHCHXSWG-2015-001,deFlorian:2016spz} and the uncertainties due to renormalisation and factorisation scale variations, parton distribution functions, and the strong coupling constant are used for the normalisation of the signal samples. In addition, uncertainties from these sources affecting the shape of the $H^{\pm\pm}$ distributions at LO are considered. An uncertainty due to the missing NLO EW corrections is adopted, as recommended in Ref.~\cite{LHCHXSWG-2015-001}.

The systematic uncertainties originating from common sources, such as those associated with the integrated luminosity, are treated as correlated between the $\hp$ and $\hpp$ channels. Uncertainties related to signal modelling as discussed above are treated as correlated.

\section{Results}
\label{sec:result}

\begin{figure}[t!bp]
\centering
\subfigure[]{
\includegraphics[width=.45\textwidth]{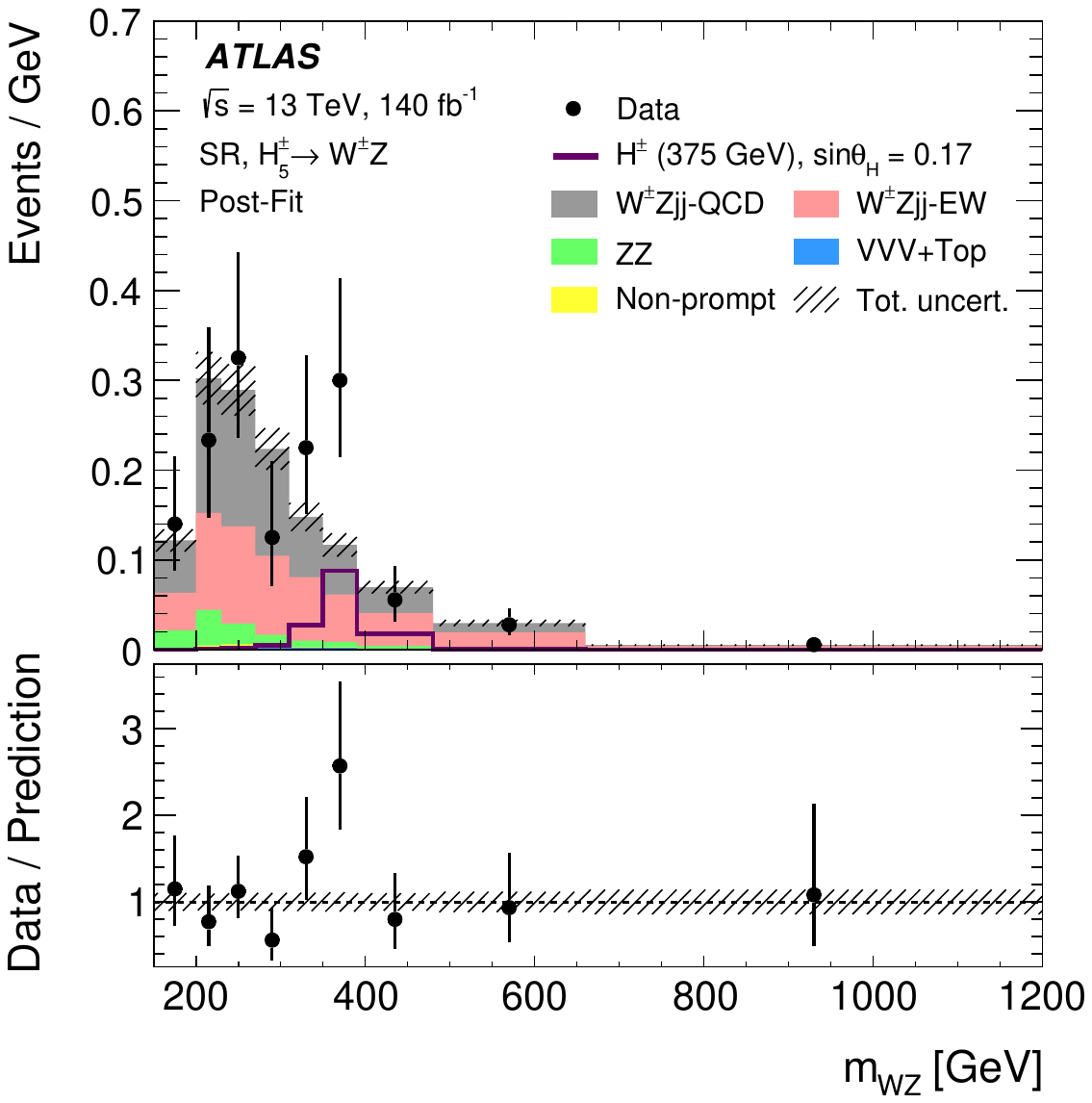}
\label{fig:results_distributions_mwz}
}
\subfigure[]{
\includegraphics[width=0.45\textwidth]{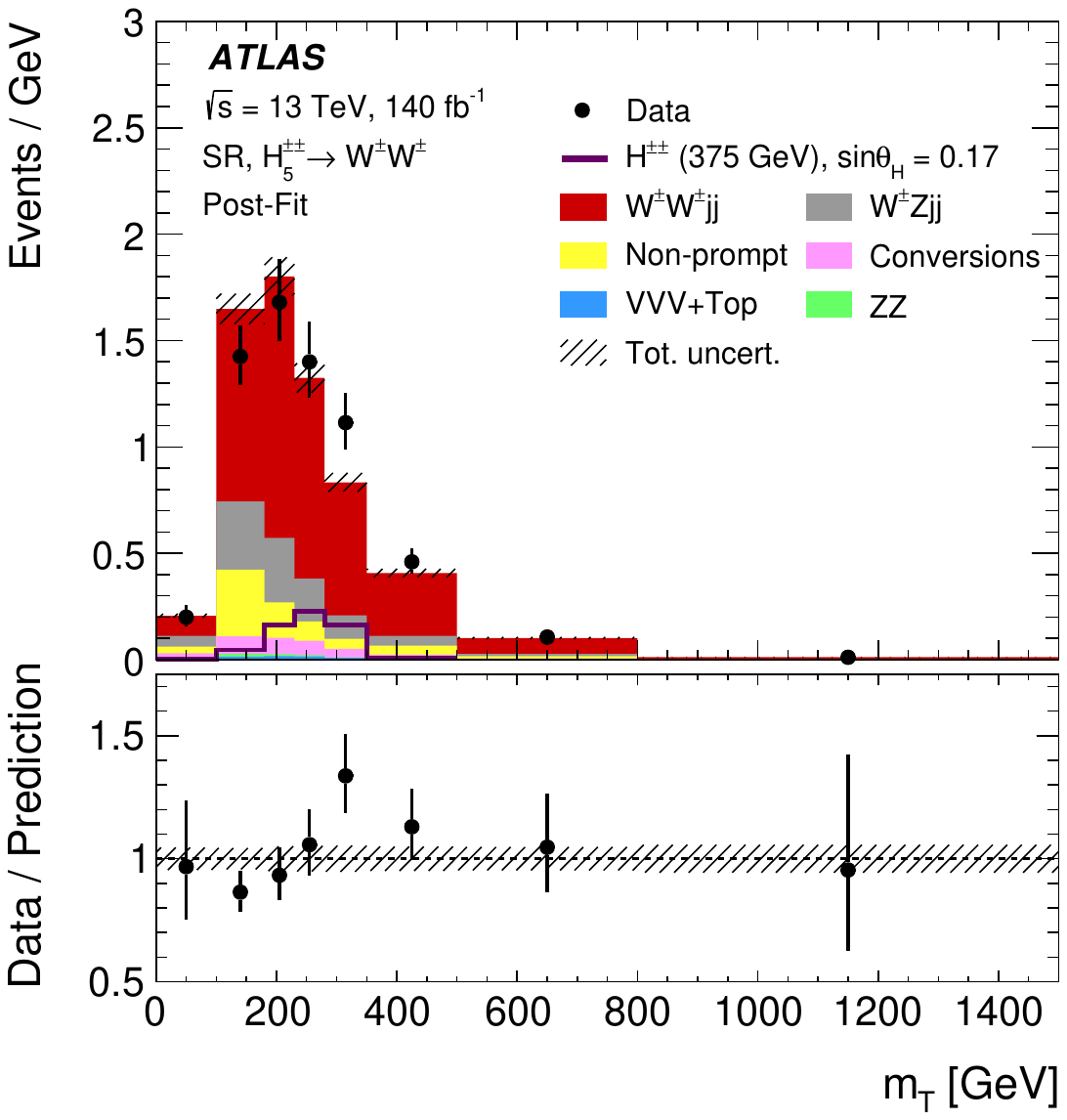}
\label{fig:results_distributions_mt}
}
\caption{The (a) $m_{\mathrm{WZ}}$ and (b) inclusive $\mt$ distributions in the signal regions for the SM background-only hypothesis. Data are shown as black markers with vertical error bars representing the statistical uncertainty. Filled histograms show contributions of various SM processes, with the hatched band representing the total uncertainty. The predicted yields are shown with their best-fit normalisations from the simultaneous fit. The last bin of each distribution includes overflow events. The lower panel shows the ratio of the data to the SM prediction. The line shows the prediction of the GM model for $\sin\theta_{\mathrm{H}} = 0.17$ and $m_{H_{5}} = 375$~\GeV, where the $\sin\theta_{\mathrm{H}}$ value corresponds to the expected 95\% CL limit for that $H_{5}$ mass.}
\label{fig:results_distributions}
\end{figure}

The distribution of $m_{\mathrm{WZ}}$ in the $\hp$ SR and the inclusive $m_{\mathrm{T}}$ distribution in the $\hpp$ SR are shown in Figure~\ref{fig:results_distributions}. The predicted yields are displayed with their best-fit normalisations obtained from the simultaneous fit under the SM background-only hypothesis. The expected contributions from a signal with $m_{H_{5}}=375~\GeV$ and $\sin\theta_{\mathrm{H}} = 0.17$ are shown for illustration purposes. The corresponding background normalisation factors for the QCD $\wzjj$, EW $\sswwjj$ and $ZZ$ processes are $0.73\pm0.06$, $1.16\pm0.11$, and $1.01\pm0.15$, respectively. These are consistent with the values reported in Refs.~\cite{atlas:2023wz,ATLAS:2023sua}. No uncertainties are significantly constrained or pulled in the simultaneous fit.

\begin{figure}[t!bp]
\centering
\subfigure[]{
\includegraphics[width=.45\textwidth]{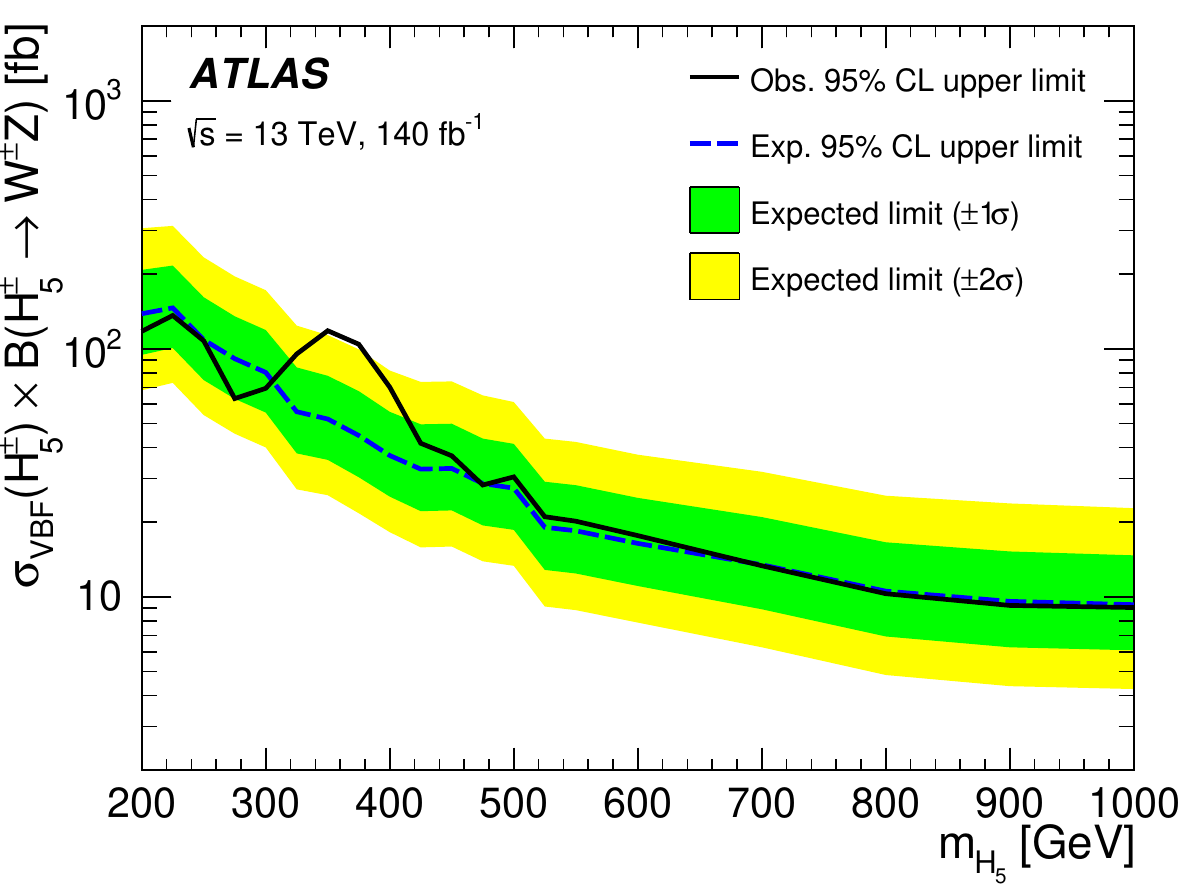}
\label{fig:results_limits_xsec_hp}
}
\subfigure[]{
\includegraphics[width=0.45\textwidth]{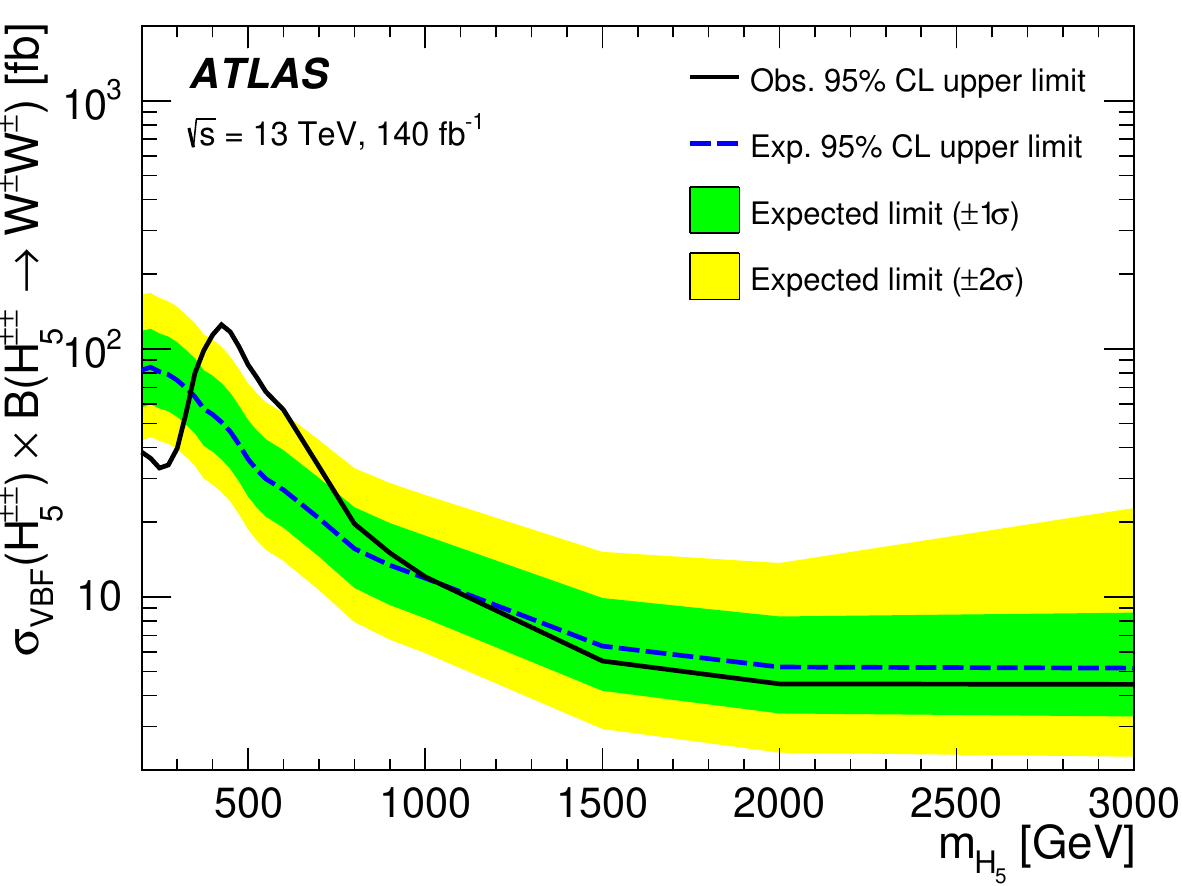}
\label{fig:results_limits_xsec_hpp}
}
\,
\subfigure[]{
\includegraphics[width=0.45\textwidth]{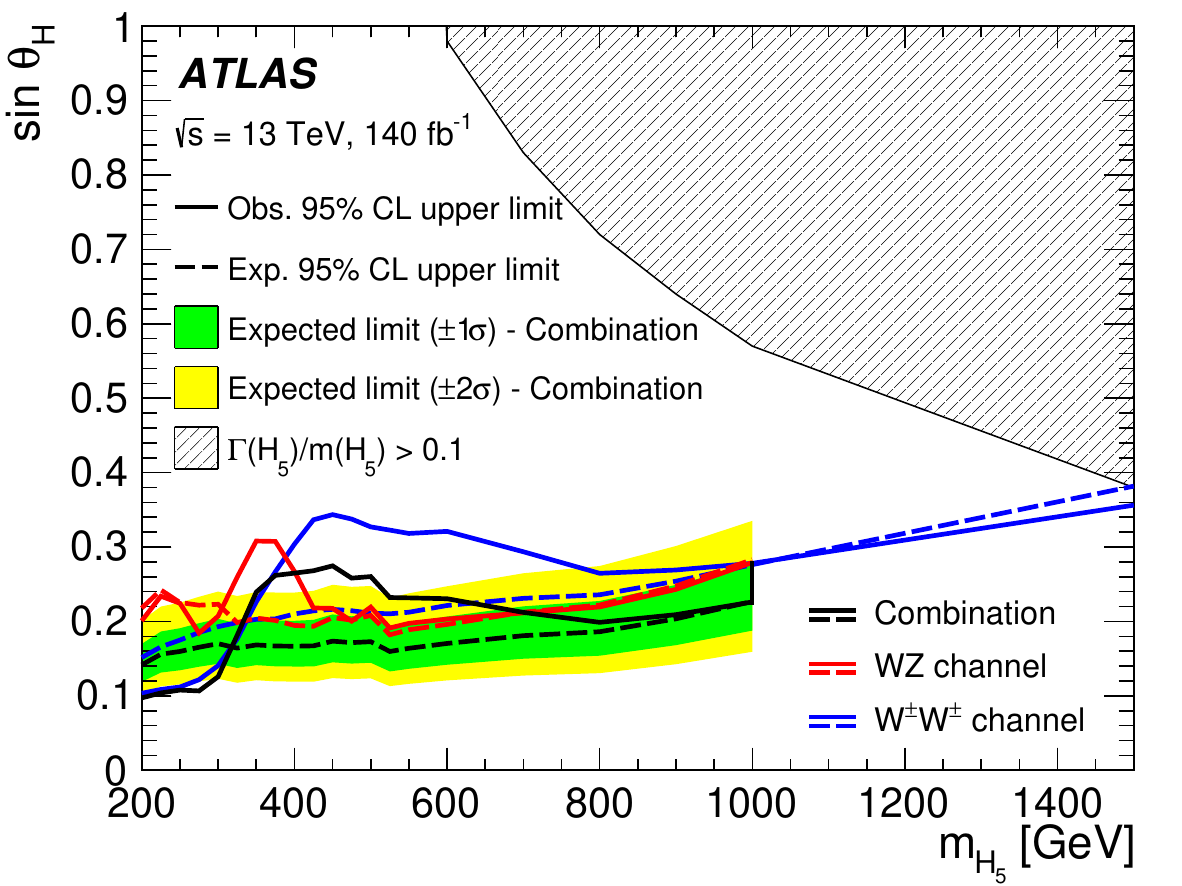}
\label{fig:results_limits_gm}
}
\caption{Expected and observed exclusion limits at 95\% CL for (a) $\sigma_{\mathrm{VBF}}(H^{\pm}_{5}) \times \mathcal{B}(\hp)$ and (b) $\sigma_{\mathrm{VBF}}(H^{\pm\pm}_{5}) \times \mathcal{B}(\hpp)$ as a function of $m_{H_{5}}$. The inner (outer) band represents the 68\% (95\%) confidence interval around the median expected limit. The exclusion limits for (c) $\sin \theta_H$ are shown up to $m_{H_{5}} = 1500$~\GeV. The limits on $\sin \theta_H$ obtained separately in the $\hp$ and $\hpp$ channels are also shown for comparison. The hatched region covers the parameter space where the intrinsic widths of the $H^{\pm}_{5}$ and $H^{\pm\pm}_{5}$ bosons would be larger than 10\% of $m_{H_{5}}$ and is disfavoured in the GM model~\cite{LHCHXSWG-2015-001}.}
\label{fig:results_limits}
\end{figure}

The 95\% CL upper limits on the production cross-section times branching fraction $\sigma_{\mathrm{VBF}}(H^{\pm}_{5}) \times \mathcal{B}(\hp)$ and $\sigma_{\mathrm{VBF}}(H^{\pm\pm}_{5}) \times \mathcal{B}(\hpp)$ for the VBF production of singly and doubly charged Higgs bosons as a function of $m_{H_{5}}$ from 200 to 3000~\GeV are shown in~\cref{fig:results_limits_xsec_hp,fig:results_limits_xsec_hpp}. The 95\% CL limits are derived using the $\text{CL}_{\textrm{s}}$ method~\cite{Junk_1999,Read:2002hq}. The asymptotic approximation~\cite{Cowan:2010js}, whose validity was confirmed through studies with pseudo-experiments, is used to derive the upper limits.

The simultaneous fit results are interpreted in the context of the GM model, setting the most stringent constraints to date on the $\sin \theta_H$ parameter as a function of $m_{H_{5}}$. The limits are shown in Figure~\ref{fig:results_limits_gm}. The black hatched region represents the parameter space for which the total width of the quintuplet states exceeds 10\% of $m_{H_{5}}$, where the model is not applicable due to considerations of perturbativity and vacuum stability, and indirect experimental constraints~\cite{LHCHXSWG-2015-001}. The combined expected limits on $\sin \theta_H$ are 10\% to 26\% more stringent, depending on $m_{H_{5}}$, than the respective limits obtained separately in the $\hp$~\cite{atlas:2023wz} and $\hpp$~\cite{ATLAS:2023sua} channels. The expected limits are 10\% to 50\% stronger, depending upon $m_{H_{5}}$, compared to the respective limits obtained by the CMS Collaboration~\cite{CMS-HIG-20-017}. The observed 95\% CL limits exclude $\sin \theta_H$ parameter values greater than 0.10--0.36 for the $m_{H_{5}}$ between 200 and 1500~\GeV.

The results show a local excess of events over the SM prediction at a resonance mass of around 400~\GeV as can be seen in Figure~\ref{fig:results_limits}. The significance of the excess has been evaluated for different $m_{H_{5}}$ in terms of the local $p$--value using the asymptotic approximation. The largest excess is for $m_{H_{5}}=375~\GeV$, with a $p$--value of $5.7 \times 10^{-4}$, corresponding to a local significance of 3.3 standard deviations. The global significance of the excess was also evaluated~\cite{Gross:2010qma}, and yields a global $p$--value of $5.6 \times 10^{-3}$, corresponding to a global significance of 2.5 standard deviations. The largest local (global) significances obtained separately from the fits performed in the $\hp$ and $\hpp$ channels used in this analysis are 2.8 (1.6) and 3.2 (2.5) for $m_{H_{5}}$ values of 375~\GeV and 450~\GeV, respectively.

\FloatBarrier

\section{Conclusion}
\label{sec:conclusion}

A combination of searches for singly and doubly charged Higgs bosons, $\hp$ and $\hpp$, produced via vector-boson fusion is reported. The dataset corresponds to 140~fb$^{-1}$ of proton--proton collision data at $\sqrt{s}= 13~\TeV$ collected with the ATLAS detector during Run 2 of the LHC (2015-2018). Constraints are reported on the production cross-section times branching fraction for singly and doubly charged Higgs bosons. The simultaneous fit results are interpreted in the context of the Georgi-Machacek model, for which the most stringent constraints to date are set. The observed 95\% CL limits exclude $\sin \theta_H$ parameter values greater than 0.10--0.36 for $m_{H_{5}}$ between 200 and 1500~\GeV. The largest deviation from the Standard Model occurs for a resonant mass near 375~\GeV, with a global significance of 2.5 standard deviations.

\section*{Acknowledgements}
%

%

%
%

%
%

We thank CERN for the very successful operation of the LHC and its injectors, as well as the support staff at
CERN and at our institutions worldwide without whom ATLAS could not be operated efficiently.

The crucial computing support from all WLCG partners is acknowledged gratefully, in particular from CERN, the ATLAS Tier-1 facilities at TRIUMF/SFU (Canada), NDGF (Denmark, Norway, Sweden), CC-IN2P3 (France), KIT/GridKA (Germany), INFN-CNAF (Italy), NL-T1 (Netherlands), PIC (Spain), RAL (UK) and BNL (USA), the Tier-2 facilities worldwide and large non-WLCG resource providers. Major contributors of computing resources are listed in Ref.~\cite{ATL-SOFT-PUB-2023-001}.

We gratefully acknowledge the support of ANPCyT, Argentina; YerPhI, Armenia; ARC, Australia; BMWFW and FWF, Austria; ANAS, Azerbaijan; CNPq and FAPESP, Brazil; NSERC, NRC and CFI, Canada; CERN; ANID, Chile; CAS, MOST and NSFC, China; Minciencias, Colombia; MEYS CR, Czech Republic; DNRF and DNSRC, Denmark; IN2P3-CNRS and CEA-DRF/IRFU, France; SRNSFG, Georgia; BMBF, HGF and MPG, Germany; GSRI, Greece; RGC and Hong Kong SAR, China; ISF and Benoziyo Center, Israel; INFN, Italy; MEXT and JSPS, Japan; CNRST, Morocco; NWO, Netherlands; RCN, Norway; MNiSW, Poland; FCT, Portugal; MNE/IFA, Romania; MSTDI, Serbia; MSSR, Slovakia; ARIS and MVZI, Slovenia; DSI/NRF, South Africa; MICIU/AEI, Spain; SRC and Wallenberg Foundation, Sweden; SERI, SNSF and Cantons of Bern and Geneva, Switzerland; NSTC, Taipei; TENMAK, T\"urkiye; STFC/UKRI, United Kingdom; DOE and NSF, United States of America.

Individual groups and members have received support from BCKDF, CANARIE, CRC and DRAC, Canada; CERN-CZ, FORTE and PRIMUS, Czech Republic; COST, ERC, ERDF, Horizon 2020, ICSC-NextGenerationEU and Marie Sk{\l}odowska-Curie Actions, European Union; Investissements d'Avenir Labex, Investissements d'Avenir Idex and ANR, France; DFG and AvH Foundation, Germany; Herakleitos, Thales and Aristeia programmes co-financed by EU-ESF and the Greek NSRF, Greece; BSF-NSF and MINERVA, Israel; NCN and NAWA, Poland; La Caixa Banking Foundation, CERCA Programme Generalitat de Catalunya and PROMETEO and GenT Programmes Generalitat Valenciana, Spain; G\"{o}ran Gustafssons Stiftelse, Sweden; The Royal Society and Leverhulme Trust, United Kingdom.

In addition, individual members wish to acknowledge support from Armenia: Yerevan Physics Institute (FAPERJ); CERN: European Organization for Nuclear Research (CERN PJAS); Chile: Agencia Nacional de Investigaci\'on y Desarrollo (FONDECYT 1230812, FONDECYT 1230987, FONDECYT 1240864); China: Chinese Ministry of Science and Technology (MOST-2023YFA1605700), National Natural Science Foundation of China (NSFC - 12175119, NSFC 12275265, NSFC-12075060); Czech Republic: Czech Science Foundation (GACR - 24-11373S), Ministry of Education Youth and Sports (FORTE CZ.02.01.01/00/22\_008/0004632), PRIMUS Research Programme (PRIMUS/21/SCI/017); EU: H2020 European Research Council (ERC - 101002463); European Union: European Research Council (ERC - 948254, ERC 101089007), Horizon 2020 Framework Programme (MUCCA - CHIST-ERA-19-XAI-00), European Union, Future Artificial Intelligence Research (FAIR-NextGenerationEU PE00000013), Italian Center for High Performance Computing, Big Data and Quantum Computing (ICSC, NextGenerationEU); France: Agence Nationale de la Recherche (ANR-20-CE31-0013, ANR-21-CE31-0013, ANR-21-CE31-0022, ANR-22-EDIR-0002), Investissements d'Avenir Labex (ANR-11-LABX-0012); Germany: Baden-Württemberg Stiftung (BW Stiftung-Postdoc Eliteprogramme), Deutsche Forschungsgemeinschaft (DFG - 469666862, DFG - CR 312/5-2); Italy: Istituto Nazionale di Fisica Nucleare (ICSC, NextGenerationEU), Ministero dell'Università e della Ricerca (PRIN - 20223N7F8K - PNRR M4.C2.1.1); Japan: Japan Society for the Promotion of Science (JSPS KAKENHI JP22H01227, JSPS KAKENHI JP22H04944, JSPS KAKENHI JP22KK0227, JSPS KAKENHI JP23KK0245); Netherlands: Netherlands Organisation for Scientific Research (NWO Veni 2020 - VI.Veni.202.179); Norway: Research Council of Norway (RCN-314472); Poland: Ministry of Science and Higher Education (IDUB AGH, POB8, D4 no 9722), Polish National Agency for Academic Exchange (PPN/PPO/2020/1/00002/U/00001), Polish National Science Centre (NCN 2021/42/E/ST2/00350, NCN OPUS nr 2022/47/B/ST2/03059, NCN UMO-2019/34/E/ST2/00393, NCN \& H2020 MSCA 945339, UMO-2020/37/B/ST2/01043, UMO-2021/40/C/ST2/00187, UMO-2022/47/O/ST2/00148, UMO-2023/49/B/ST2/04085, UMO-2023/51/B/ST2/00920); Slovenia: Slovenian Research Agency (ARIS grant J1-3010); Spain: Generalitat Valenciana (Artemisa, FEDER, IDIFEDER/2018/048), Ministry of Science and Innovation (MCIN \& NextGenEU PCI2022-135018-2, MICIN \& FEDER PID2021-125273NB, RYC2019-028510-I, RYC2020-030254-I, RYC2021-031273-I, RYC2022-038164-I), PROMETEO and GenT Programmes Generalitat Valenciana (CIDEGENT/2019/027); Sweden: Carl Trygger Foundation (Carl Trygger Foundation CTS 22:2312), Swedish Research Council (Swedish Research Council 2023-04654, VR 2018-00482, VR 2022-03845, VR 2022-04683, VR 2023-03403, VR grant 2021-03651), Knut and Alice Wallenberg Foundation (KAW 2018.0157, KAW 2018.0458, KAW 2019.0447, KAW 2022.0358); Switzerland: Swiss National Science Foundation (SNSF - PCEFP2\_194658); United Kingdom: Leverhulme Trust (Leverhulme Trust RPG-2020-004), Royal Society (NIF-R1-231091); United States of America: U.S. Department of Energy (ECA DE-AC02-76SF00515), Neubauer Family Foundation.

%
%


%
%
%
%
%
%

%
%

%
%
%
\printbibliography
\clearpage
 
\begin{flushleft}
\hypersetup{urlcolor=black}
{\Large The ATLAS Collaboration}

\bigskip

\AtlasOrcid[0000-0002-6665-4934]{G.~Aad}$^\textrm{\scriptsize 104}$,
\AtlasOrcid[0000-0001-7616-1554]{E.~Aakvaag}$^\textrm{\scriptsize 17}$,
\AtlasOrcid[0000-0002-5888-2734]{B.~Abbott}$^\textrm{\scriptsize 123}$,
\AtlasOrcid[0000-0002-0287-5869]{S.~Abdelhameed}$^\textrm{\scriptsize 119a}$,
\AtlasOrcid[0000-0002-1002-1652]{K.~Abeling}$^\textrm{\scriptsize 56}$,
\AtlasOrcid[0000-0001-5763-2760]{N.J.~Abicht}$^\textrm{\scriptsize 50}$,
\AtlasOrcid[0000-0002-8496-9294]{S.H.~Abidi}$^\textrm{\scriptsize 30}$,
\AtlasOrcid[0009-0003-6578-220X]{M.~Aboelela}$^\textrm{\scriptsize 45}$,
\AtlasOrcid[0000-0002-9987-2292]{A.~Aboulhorma}$^\textrm{\scriptsize 36e}$,
\AtlasOrcid[0000-0001-5329-6640]{H.~Abramowicz}$^\textrm{\scriptsize 155}$
% ...
\bigskip
\\

% ...
$^{17}$Department for Physics and Technology, University of Bergen, Bergen; Norway.\\
% ...
$^{30}$Physics Department, Brookhaven National Laboratory, Upton NY; United States of America.\\
% ...
$^{36}$$^{(a)}$Facult\'e des Sciences Ain Chock, Universit\'e Hassan II de Casablanca;$^{(b)}$Facult\'{e} des Sciences, Universit\'{e} Ibn-Tofail, K\'{e}nitra;$^{(c)}$Facult\'e des Sciences Semlalia, Universit\'e Cadi Ayyad, LPHEA-Marrakech;$^{(d)}$LPMR, Facult\'e des Sciences, Universit\'e Mohamed Premier, Oujda;$^{(e)}$Facult\'e des sciences, Universit\'e Mohammed V, Rabat;$^{(f)}$Institute of Applied Physics, Mohammed VI Polytechnic University, Ben Guerir; Morocco.\\
% ...
$^{45}$Physics Department, Southern Methodist University, Dallas TX; United States of America.\\
% ...
$^{50}$Fakult\"{a}t Physik , Technische Universit{\"a}t Dortmund, Dortmund; Germany.\\
% ...
$^{56}$II. Physikalisches Institut, Georg-August-Universit\"{a}t G\"ottingen, G\"ottingen; Germany.\\
% ...
$^{104}$CPPM, Aix-Marseille Universit\'e, CNRS/IN2P3, Marseille; France.\\
% ...
$^{119}$$^{(a)}$New York University Abu Dhabi, Abu Dhabi;$^{(b)}$United Arab Emirates University, Al Ain; United Arab Emirates.\\
% ...
$^{123}$Homer L. Dodge Department of Physics and Astronomy, University of Oklahoma, Norman OK; United States of America.\\
% ...
$^{155}$Raymond and Beverly Sackler School of Physics and Astronomy, Tel Aviv University, Tel Aviv; Israel.\\
% ...
$^{a}$ Also Affiliated with an institute covered by a cooperation agreement with CERN.\\
$^{b}$ Also at An-Najah National University, Nablus; Palestine.\\
$^{c}$ Also at Borough of Manhattan Community College, City University of New York, New York NY; United States of America.\\
$^{d}$ Also at Center for Interdisciplinary Research and Innovation (CIRI-AUTH), Thessaloniki; Greece.\\
$^{e}$ Also at CERN, Geneva; Switzerland.\\
$^{f}$ Also at CMD-AC UNEC Research Center, Azerbaijan State University of Economics (UNEC); Azerbaijan.\\
% ...

\end{flushleft}


%
\FloatBarrier

\end{document}